\documentclass[aps,showpacs,twocolumn,superscriptaddress]{revtex4}
\usepackage{graphicx}
\usepackage{dcolumn}
\begin{document}

\title{Traces of $\Theta^+$ pentaquark in $K^+$ - nucleus
dynamics}

\author{A.~Gal}
\affiliation{Racah Institute of Physics, The Hebrew University,
Jerusalem 91904, Israel\vspace*{1ex}}

\author{E.~Friedman}
\affiliation{Racah Institute of Physics, The Hebrew University,
Jerusalem 91904, Israel\vspace*{1ex}}

\begin{abstract}
\rule{0ex}{3ex}

Long-standing anomalies in $K^+$ - nucleus integral cross sections
could be resolved by extending the impulse-approximation $t\rho$
optical-potential framework to incorporate $K^+$ absorption on pairs
of nucleons. Substantially improved fits to the data at $p_{\rm lab}
\sim 500 - 700$ MeV/c are obtained. An upper bound on the absorption
cross section per nucleon is derived, $\sigma_{\rm abs}^{(K^+)}/A
\sim 3.5$ mb. We conjecture that the underlying microscopic absorption
process is $K^+ nN \rightarrow \Theta^+ N$, where $\Theta^+$(1540) is
the newly discovered exotic $Y$=2, $I$=0, $Z$=1 pentaquark baryon,
and estimate that $\sigma (K^+ d \rightarrow \Theta^+ p)$ is a fraction
of millibarn. Comments are made on $\Theta^+$ production reactions on 
nuclei. 
\end{abstract}

\maketitle

%{\it Keywords}: $K^+$ - nucleus reaction and total cross sections;
%$K^+ NN$ absorption; $\sigma (K^+ d \rightarrow \Theta^+ p)$;
%$\Theta^+$(1540) pentaquark nuclear interactions.

\section{Introduction and motivation}
\label{sec:int}

The $K^+N$ interaction below the pion-production threshold is fairly weak and
featureless, as anticipated from an `exotic' channel corresponding to quark
content $qqqq{\bar s}$, where $q$ denotes a light nonstrange quark.
This merit has motivated past suggestions to probe nuclear in-medium effects
by studying scattering and reaction processes with $K^+$ beams below 800
MeV/c; see Ref.\cite{DWa82} for an early review.
Limited total cross-section data \cite{BGK68} on carbon, and elastic
and inelastic differential cross section data \cite{MBC82} on carbon and
calcium, drew theoretical attention already in the 1980s to the insufficiency
of the impulse-approximation $t\rho$ form of the $K^+$ - nucleus optical
potential, where $t$ is the free-space $K^+N$ $t$ matrix, particularly
with respect to its reaction content (`reactivity' below). In order to
account for the increased reactivity in $K^+$ - nucleus interactions,
Siegel {\it et al.}\cite{SKG85} suggested that nucleons `swell' in the
nuclear medium, primarily by increasing the dominant hard-core $S_{11}$
phase shift. Brown {\it et al.}\cite{BDS88} suggested that the extra
reactivity was due to the reduced in-medium masses of exchanged vector
mesons, and this was subsequently worked out in detail in Ref.\cite{CLa96}.
Another source for increased reactivity in $K^+$ - nucleus interactions
was discussed in the 1990s and is due to meson exchange-current effects
\cite{JKo92,GNO95}.

Some further experimental progress was made during the early 1990s,
consisting mostly of measuring attenuation cross sections in $K^+$
transmission experiments at the BNL-AGS on deuterium and several other
nuclear targets in the momentum range $p_{\rm lab} = 450 - 740$ MeV/c
\cite{KAA92,SWA93,WAA94,FGW97} and of measuring $K^+$ quasielastic
scattering on several targets at 705 MeV/c \cite{KPS95}. New measurements
of $K^+$ elastic and inelastic differential cross sections on C and Ca at
715 MeV/c were reported in Ref.\cite{MBB96} and analyzed in Ref.\cite{CSP97},
and self-consistent final values of $K^+$ integral (reaction and total) cross
sections on $^6$Li, $^{12}$C, $^{28}$Si and $^{40}$Ca were published in
Refs.\cite{FGM97a,FGM97b}. By the late 1990s, experimentation in
$K^+$ - nuclear physics has subsided, and with it died out also theoretical
interest. The subject was reviewed last time in HYP97,
concluding that ``every experiment of $K^+$ mesons with complex nuclei finds
cross sections larger than those predicted" \cite{Pet98}; and based on
analyses of $K^+$ - nuclear integral cross sections \cite{FGM97a,FGM97b},
it was concluded that ``at present theory misses some unconventional in-medium
effects" \cite{Gal98}.

The $\Theta^+$(1540) exotic baryon \cite{PDG04} provides a new mode of
reactivity to $K^+$ - nuclear interactions.
The $\Theta^+$ couples directly to nucleons,
$\Theta^+ \rightarrow K^+n,K^0p$, but its small width of order 1 MeV or
less \cite{CTr04,Gib04,SHK04} indicates that this coupling is weak. The
$K^+n$ phase-shift input to the $t\rho$ optical potential in the vicinity
of the $\Theta^+$ mass (corresponding to $p_{\rm lab} \sim 440$ MeV/c)
is derived from $K^+d$ scattering data across the $\Theta^+$ resonance
energy. Therefore, whatever the $KN\Theta$ coupling is, its effect is
already included at least implicitly in the $t\rho$ impulse approximation.
A deuteron target also allows for a two-body production reaction
$K^+d \rightarrow \Theta^+p$ with threshold at $p_{\rm lab} \sim 400$
MeV/c \cite{Ash04}, but the available $K^+d$ scattering data do not give
any evidence for the opening of such a production channel.
It is plausible that {\it denser} nuclear targets would help
to fuse the $K^+$'s $\bar s$ quark together with two
$(ud)_{S=0,I=0}$ diquarks, each of which belongs to a distinct nucleon,
into the $(ud)(ud)\bar s$ configuration according to the Jaffe-Wilczek
model of $\Theta^+$ \cite{JWi03}, and similarly for the diquark-triquark
Karliner-Lipkin model \cite{KLi03}. Therefore, one should look for traces
of $K^+$ {\it absorption} on two nucleons, $K^+nN \rightarrow \Theta^+N$,
in normal nuclei. In the present work we show that traces of such two-body
$K^+$ absorption may be identified in $K^+$ - nucleus dynamics and help
resolve the long-standing anomalies in the $K^+$ - nucleus integral cross
sections. We determine the $K^+$ absorption cross section on nuclei at 
$p_{\rm lab}=488$ MeV/c which is the closest momentum to the $\Theta^+$ 
rest mass, where good data are available. 
%In addition, we comment briefly on the implications of such
%traces on the $\Theta^+$ - nucleus interaction in connection 
%with several recent suggestions that this interaction may be
%sufficiently strong to bind $\Theta^+$ in nuclei \cite{Mil04,CLM04}.
Our results provide the first concrete demonstration that nuclear targets
are potentially useful in the study of exotic baryons.

\section{Methodology, Results and Discussion}
\label{sec:res}

\begin{figure}[t]
\centerline{\includegraphics[height=6.8cm]{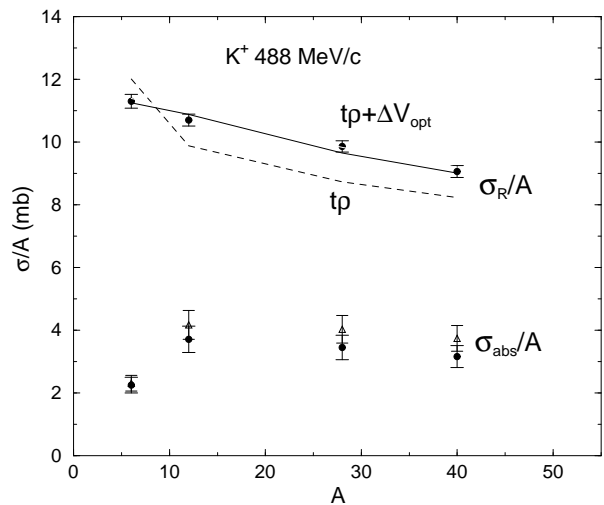}}
\caption{Data and calculations for $K^+$ reaction cross sections per
nucleon ($\sigma_R/A$) at $p_{\rm lab}=488$ MeV/c are shown in the
upper part. Calculated $K^+$ absorption cross sections per nucleon
($\sigma_{\rm abs}/A$) are shown in the lower part, see text.}
\label{kplusfig1}
\end{figure}

In the calculations presented below the Klein Gordon equation is solved,
using the simplest possible $t\rho$ form for the optical potential
\begin{equation}
\label{equ:Vopt}
2 \varepsilon^{(A)}_{\rm red} V_{\rm opt}(r) = -4\pi F_A b_0 \rho(r) ~~,
\end{equation}
where $\varepsilon^{(A)}_{\rm red}$ is the reduced energy in the cm
system, $F_A$ is a kinematical factor resulting from the transformation
of amplitudes between the $KN$ and the $K^+$ - nucleus cm systems and
$b_0$ is the (complex) value of the isospin-averaged $KN$ scattering
amplitude in the forward direction. The Coulomb potential due to the
charge distribution of the nucleus is included.
This form of the potential takes into account $1/A$ corrections,
an important issue when handling as light a nucleus as $^6$Li.
Using this approach Friedman {\it et al.}\cite{FGM97a} showed
that no {\it effective} value for $b_0$ could be found that fits
satisfactorily the reaction and total cross sections derived from the
BNL-AGS transmission measurements at $p_{\rm lab} = 488, 531, 656, 714$
MeV/c on $^6$Li, $^{12}$C, $^{28}$Si, $^{40}$Ca. This is demonstrated
in the upper part of Fig. \ref{kplusfig1} for the reaction cross sections
per nucleon $\sigma_R/A$ at 488 MeV/c, where the calculated cross sections 
using a best-fit $t\rho$ optical potential (dashed line) are compared 
with the experimental values listed in Ref. \cite{FGM97b}. 
The best-fit values of 
Re$b_0$ and Im$b_0$ which specify this $t\rho$ potential are given in the 
first row of Table \ref{tab:FGa04} where Im$b_0$ represents $10-15\%$ 
increase with respect to the free-space value. The $\chi ^2$/N of this 
density-independent fit is very high. Its failure is due to the 
impossibility to reconcile the $^6$Li data (which for the total cross 
sections are consistent with the $K^+d$ `elementary' cross sections) 
with the data on the other, denser nuclei, as is clearly exhibited in 
Fig. \ref{kplusfig1} for the best-fit $t\rho$ dashed line. 
If $^6$Li were removed out of the data base, then it would have become 
possible to fit reasonably well the rest of the nuclei, but the rise 
in Im$b_0$ would then be substantially higher than that for the $t\rho$ 
potential used here. Such fits, excluding $^6$Li, are less successful at 
the higher energies. 

To incorporate $K^+nN \rightarrow \Theta^+N$ two-nucleon absorption
into the impulse-approximation motivated $V_{\rm opt}(r)$,
Eq.(\ref{equ:Vopt}), we add a $\rho^2 (r)$ piece, as successfully
practised in pionic atoms to account for $\pi^-$ absorption on two
nucleons:
\begin{equation}
\label{equ:DD1}
b_0~ \rho (r) \rightarrow b_0~ \rho (r)~+~B~ \rho^2 (r)~, 
\end{equation} 
where the parameter $B$ reperesents the effect of $K^+$ nuclear absorption 
into exotic $S=+1$ baryonic channels. Using this potential we have repeated 
fits to all 32 data points for the reaction and total cross sections.
As there were correlations between the parameters of Re$V_{\rm opt}$ 
we subsequently kept the parameter Re$b_0$ fixed at its free kaon-nucleon 
value. The results are summarized in Table \ref{tab:FGa04} (marked as 
`Eq.(\ref{equ:DD1})') and the substantial improvement compared to the 
$t\rho$ potential is self evident. However, the fits at the higher
momenta are not as successful as the fit at 488 MeV/c, suggesting that
one needs a more effective way to distinguish between $^6$Li and the denser
nuclear targets. In fact, it was shown {\it empirically} \cite{FGM97a,FGM97b} 
that the average nuclear density
${\bar \rho}=\frac{1}{A}\int\rho^2d{\bf r}$ provides such discrimination 
and is instrumental in achieving good agreement with experiment. Therefore, 
we replace Eq.(\ref{equ:DD1}) by the simplest ansatz 
\begin{equation}
\label{equ:DD2}
b_0~ \rho (r) \rightarrow b_0~ \rho (r)~+~B~ {\bar \rho}~\rho (r)~. 
\end{equation} 
The added piece is a functional of the density which to lowest order 
reduces to a $\rho^2$ form. To justify it theoretically would require 
to understand quantitatively the production mechanisms of the $\Theta^+$ 
pentaquark in dense nuclear matter. The resulting fits are shown in 
Table \ref{tab:FGa04}, marked as `Eq.(\ref{equ:DD2})'.
The superiority of this ${\bar \rho} \rho$ version compared to the
$\rho ^2$ one is very clearly observed.
This conclusion remains valid also when the data for the elastic scattering
of 715 MeV/c $K^+$ by $^6$Li and $^{12}$C \cite{MBB96} are included
in the analysis. The calculated reaction cross sections at 488 MeV/c, 
using Eq.(\ref{equ:DD2}), are shown by the solid line marked 
$t\rho + \Delta V_{\rm opt}$ in the upper part of Fig. \ref{kplusfig1}, 
where $\Delta V_{\rm opt}$ is the added piece of $V_{\rm opt}$ due to 
a nonzero value of $B$. Clearly, it is a very good fit. 

Inspecting the results in Table \ref{tab:FGa04} one notes that
the splitting of Im$V_{\rm opt}$ into its two reactive components Im$b_0$
and Im$B$ appears well determined by the data at all energies, with very
accurate values of Im$b_0$ thus derived. These values of Im$b_0$ are close
to, but somewhat below the corresponding free-space values. The two-nucleon
absorption coefficient Im$B$ rises slowly with energy as appropriate to the
increased phase space available to the underlying two-nucleon absorption
process $K^+nN \rightarrow \Theta^+N$. Its values in this energy range 
are roughly independent of the form of $\Delta V_{\rm opt}$, the more 
conservative Eq.(\ref{equ:DD1}) or the more effective Eq.(\ref{equ:DD2}), 
used to derive these values from the data. 
Regarding Re$V_{\rm opt}$, and recalling that 
${\bar \rho} \sim 0.1~ {\rm fm}^{-3}$ for the dense nuclear targets, 
it is clear that Re$V_{\rm opt} \sim 0$ at the two higher momenta, 
illustrating the inadequacy of the $t\rho$ model in which the best-fit 
solution flips from repulsion at the two lower momenta (consistently 
with the impulse approximation) to attraction of a similar order of 
magnitude at the two higher momenta. 
We note that our $K^+nN \rightarrow \Theta^+N$ absorption reaction 
is related to the mechanism proposed recently in Ref. \cite{CLM04} 
as causing strong $\Theta^+$ - nuclear attraction, based on $K\pi$ two-meson 
cloud contributions to the self energy of $\Theta^+$ in nuclear matter. 
However, it would appear difficult to reconcile as strong 
$\Theta^+$ - nuclear attraction as proposed there with the magnitude of 
Re$B$ derived in the present work.  

\begin{table}
\caption{Fits to the eight $K^+$-nuclear integral cross sections at four
laboratory momenta $p_{\rm lab}$ (in MeV/c), using different potentials.}
\label{tab:FGa04}
\begin{tabular}{ccccccc}
\hline \hline
$p_{\rm lab}$&$V_{\rm opt}$&Re$b_0$(fm)&Im$b_0$(fm)&Re$B$(fm$^4$)&
Im$B$(fm$^4$)&$\chi ^2$/N  \\  \hline
488&$t\rho$ &$-$0.205(27)&0.173(7)& & & 18.2 \\
   &Eq.(\ref{equ:DD1}) &$-$0.178&0.120(6)&0.80(34)&0.92(8) &1.60 \\
   &Eq.(\ref{equ:DD2}) &$-$0.178&0.126(4)&0.19(11)&0.67(6)&0.69 \\
 & & & & & & \\
531&$t\rho$ &$-$0.198(41)&0.203(10)& & & 63.4 \\
   &Eq.(\ref{equ:DD1}) &$-$0.172&0.157(13)&1.80(35)&0.68(24) &7.38 \\
   &Eq.(\ref{equ:DD2}) &$-$0.172&0.144(7)&0.50(28)&0.82(9)&6.06 \\
 & & & & & & \\
656&$t\rho$ &0.168(34)&0.250(11)& & & 38.2 \\
   &Eq.(\ref{equ:DD1}) &$-$0.165&0.205(17)&1.75(55)&0.85(31) &8.78 \\
   &Eq.(\ref{equ:DD2}) &$-$0.165&0.203(5)&2.16(19)&0.78(8)&1.42 \\
 & & & & & & \\
714&$t\rho$ &0.176(39)&0.275(13)& & & 51.0 \\
   &Eq.(\ref{equ:DD1}) &$-$0.161&0.221(21)&1.57(83)&1.04(40) &11.4 \\
   &Eq.(\ref{equ:DD2}) &$-$0.161&0.218(8)&1.75(41)&0.97(12)&2.40 \\
\hline \hline
\end{tabular}
\end{table}

The $K^+$ nuclear absorption cross section $\sigma_{\rm abs}^{(K^+)}$
due to the availability of $\Theta^+$ - nuclear final states is driven
by Im($\Delta V_{\rm opt}$). We approximate it by 
\begin{equation}
\label{equ:abs}
\sigma_{\rm abs}^{(K^+)} \sim  -~ {\frac{2}{\hbar v}}
\int {\rm Im} (\Delta V_{\rm opt}(r))~
|\Psi^{(+)}({\bf r})|^2~d{\bf r} ~~,
\end{equation}
where the distorted waves $\Psi^{(+)}$ are evaluated 
in two different ways in order to assess the theoretical uncertainty. 
Calculated absorption cross sections {\it per target nucleon} at
$p_{\rm lab}=488$ MeV/c are shown in the lower part of Fig. \ref{kplusfig1}
for the fit using Eq. (\ref{equ:DD2}) for $V_{\rm opt}$ in Table 
\ref{tab:FGa04}. The triangles are for the case where $\Delta V_{\rm opt}$ 
does not enter the evaluation of the distorted waves $\Psi^{(+)}$, 
as appropriate to the DWIA approximation, and the solid circles are 
for the case where $\Psi^{(+)}$ are the fully distorted waves, including 
the effect of $\Delta V_{\rm opt}(r)$. 
The error bars plotted are due to the uncertainty in the
parameter Im$B$. It is seen that these calculated absorption cross sections,
for the relatively dense targets of C, Si and Ca, are proportional to the
mass number $A$, and the cross section per target nucleon due to
Im$B \neq 0$ is estimated as close to 3.5 mb. Although the less successful 
Eq.(\ref{equ:DD1}) gives cross sections larger by $40\%$, this value 
should be regarded an upper limit, since the best-fit density-dependent 
potentials of Refs.\cite{FGM97a,FGM97b} yield values smaller than 3.5 mb 
by a similar amount. 
The experience gained from studying $\pi$-nuclear absorption \cite{ASc86}
leads to the conclusion that $\sigma_{\rm abs}(K^+NN)$ is smaller than the
extrapolation of $\sigma_{\rm abs}^{(K^+)}/A$ in Fig. \ref{kplusfig1} to
$A=1$, and since the $KN$ interaction is weaker than the $\pi N$ interaction
one expects a reduction of roughly $50\%$, so that
$\sigma_{\rm abs}(K^+NN) \sim 1 - 2$ mb.

We note in Fig.\ref{kplusfig1} the considerably smaller absorption cross
section per nucleon calculated for $^6$Li which, considering its low density,
suggests a cross section of order fraction of millibarn for
$K^+ d \rightarrow \Theta^+ p$, well below the order 1 mb which as Gibbs has
argued recently \cite{Gib04} could indicate traces of $\Theta^+$ in $K^+ d$
total cross sections near $p_{\rm lab} \sim 440$ MeV/c. To be definite, 
we assume that $\sigma(K^+ d \rightarrow \Theta^+ p) \sim 0.1~-~0.5$ mb. 
It is worth noting that this {\it two-nucleon} cross section is considerably 
larger than what a {\it one-nucleon} production process 
$K^+ n \rightarrow \Theta^+$ would induce on a deuteron target.
Such a one-step process at the $\Theta^+$ mass ($p_{\rm lab}=440$ MeV/c)
\begin{equation}
\label{equ:Theta}
K^+ ~n \rightarrow \Theta^+ ~, ~~~ \Theta^+ ~N \rightarrow \Theta^+ ~N
\end{equation}
may be compared to pion absorption near the (3,3) resonance energy
\cite{ASc86}, where the primary two-nucleon absorption mechanism is
through a direct $\Delta$ production
\begin{equation}
\label{equ:Delta}
\pi^+ ~p \rightarrow \Delta^{++}~, ~~~ \Delta^{++}~n \rightarrow p~p ~,
\end{equation}
with $\sigma(\pi^+ d \rightarrow pp) \sim 12.5$ mb.
Scaling this value by the ratio of coupling constants squared
$g^2_{KN\Theta}/g^2_{\pi N \Delta} \sim 2.5 \times 10^{-3}$, assuming
$J^{\pi}(\Theta^+) = {(\frac{1}{2})}^+$ and
$\Gamma (\Theta^+ \rightarrow K N) \sim 1$ MeV, we estimate a cross section
level of 0.03 mb for the one-step production process at the $\Theta^+$
resonance energy. [Assuming $J^{\pi}(\Theta^+) = {(\frac{1}{2})}^-$, the
one-step production cross section is lower by at least another order of
magnitude.] The one-step cross section affordable by the neutron Fermi
motion at $p_{\rm lab}=488$ MeV/c would be considerably smaller than this
estimate \cite{Lip04}. In contrast, the two-nucleon reaction need not
involve the suppressed $KN\Theta$ coupling and its cross section,
$\sigma(K^+ d \rightarrow \Theta^+ p) \sim 0.1~-~0.5$ mb at
$p_{\rm lab}=488$ MeV/c, should depend smoothly on energy.

\section{Conclusions}
\label{sec:conc}

In conclusion, we have demonstrated a very 
good agreement between experiment and calculation for all the available 
integral cross-section data by adding to the $t\rho$ optical potential 
a density-dependent term which simulates absorption channels. 
We have identified these absorption channels, as exhibited by the 
anomalous reactivity established systematically in past phenomenological 
analyses of $K^+$ - nuclear interactions, with the $\Theta^+$ production 
reaction $K^+ nN \rightarrow \Theta^+ N$ with threshold at 
$p_{\rm lab} \sim 400$ MeV/c. The analysis of these 
data is consistent with an upper limit of about 3.5 mb on the $K^+$ 
absorption cross section per nucleon, for $\Theta^+$ production on the 
denser nuclei of C, Si, Ca, and indicates a sub-millibarn cross section 
for $\Theta^+$ production on deuterium.
We urge experimenters to look for the $K^+ d \rightarrow \Theta^+ p$
two-body production reaction \cite{Ash04} which requires experimental
accuracies of 0.1 mb in cross section measurements. Given the magnitude 
of the $K^+$ nuclear absorption cross sections, as derived in the present 
work, we also urge doing ($K^+,p$) experiments on nuclear targets. This 
reaction which has a `magic momentum' about $p_{\rm lab} \sim 600$ MeV/c, 
where the $\Theta^+$ is produced at rest, is particularly suited to study 
bound or continuum states in {\it hyponuclei} \cite{Gol82}. It might prove 
more useful than the large momentum transfer ($K^+,\pi^+$) reaction 
proposed in this context \cite{NHO04}. 
Undoubtedly, precise $K^+$ - nuclear scattering and reaction data would be
extremely useful to obtain further, more direct evidence for the presence
of the $\Theta^+$ exotic baryon and its effects in the nuclear medium.
In particular, $K^+d$ and $K^+$ - nuclear data in the range
$p_{\rm lab} \sim 300-500$ MeV/c would be very helpful to study the onset
of strange-pentaquark dynamics in the nuclear medium.

\begin{acknowledgments}

This work was supported in part by the Israel Science Foundation grant 131/01.

\end{acknowledgments}

\end{document}